\documentclass[twocolumn,fleqn,usenatbib,linenumber]{aastex63}

\usepackage{mathptmx}
\usepackage[T1]{fontenc}
\usepackage{ae,aecompl}

\usepackage{graphicx}   
\usepackage{amsmath}    
\usepackage{amssymb}    
\DeclareMathOperator{\sech}{sech}

\received{XX XXXX}
\revised{XX XXXX}
\accepted{XX XXXX}
\submitjournal{ApJ}

\shorttitle{Nova Rate with 3D Dust Maps}
\shortauthors{Kawash et al.}

\begin{document}

\title{Galactic Extinction: How Many Novae Does it Hide and How Does it Affect the Galactic Nova Rate?}

\correspondingauthor{Adam Kawash}
\email{kawashad@msu.edu}

\author[0000-0003-0071-1622]{A.\ Kawash}
\affiliation{Center for Data Intensive and Time Domain Astronomy, Department of Physics and Astronomy, Michigan State University, East Lansing, MI 48824, USA}

\author{L.\ Chomiuk}
\affiliation{Center for Data Intensive and Time Domain Astronomy, Department of Physics and Astronomy, Michigan State University, East Lansing, MI 48824, USA}

\author{J. A. \ Rodriguez}
\affiliation{Center for Data Intensive and Time Domain Astronomy, Department of Physics and Astronomy, Michigan State University, East Lansing, MI 48824, USA}

\author{J.\ Strader}
\affiliation{Center for Data Intensive and Time Domain Astronomy, Department of Physics and Astronomy, Michigan State University, East Lansing, MI 48824, USA}

\author[0000-0001-5991-6863]{K.~V.\ Sokolovsky}
\affiliation{Center for Data Intensive and Time Domain Astronomy, Department of Physics and Astronomy, Michigan State University, East Lansing, MI 48824, USA}
\affiliation{Sternberg Astronomical Institute, Moscow State University, Universitetskii~pr.~13, 119992~Moscow, Russia}

\author{E.\ Aydi}
\affiliation{Center for Data Intensive and Time Domain Astronomy, Department of Physics and Astronomy, Michigan State University, East Lansing, MI 48824, USA}

\author{C. S. \ Kochanek}
\affiliation{Department of Astronomy, The Ohio State University, 140 West 18th Avenue, Columbus, OH 43210, USA}
\affiliation{Center for Cosmology and Astroparticle Physics, The Ohio State University, 191 W. Woodruff Avenue, Columbus, OH 43210, USA}

\author{K.\ Z.\ Stanek}
\affiliation{Department of Astronomy, The Ohio State University, 140 West 18th Avenue, Columbus, OH 43210, USA}
\affiliation{Center for Cosmology and Astroparticle Physics, The Ohio State University, 191 W. Woodruff Avenue, Columbus, OH 43210, USA}

\author{K.\ Mukai}
\affiliation{CRESST II and X-ray Astrophysics Laboratory, NASA/GSFC, Greenbelt, MD 20771, USA}
\affiliation{Department of Physics, University of Maryland, Baltimore County, 1000 Hilltop Circle, Baltimore, MD 21250, USA}

\author{K.\ De}
\affiliation{Cahill Center for Astrophysics, California Institute of Technology, 1200 E. California Blvd. Pasadena, CA 91125, USA.}

\author{B.\ Shappee}
\affiliation{Institute for Astronomy, University of Hawai`i at M\=anoa, 2680 Woodlawn Dr., Honolulu 96822, USA}

\author{ T.~W.-S.\ Holoien}
\affiliation{Carnegie Observatories, 813 Santa Barbara Street, Pasadena, CA 91101, USA}

\author{J.~L.\ Prieto}
\affiliation{N\'ucleo de Astronom\'ia de la Facultad de Ingenier\'ia y Ciencias, Universidad Diego Portales, Av. Ej\'ercito 441, Santiago, Chile}
\affiliation{Millennium Institute of Astrophysics, Santiago, Chile }

\author{T. A. Thompson}
\affiliation{Department of Astronomy, The Ohio State University, 140 West 18th Avenue, Columbus, OH 43210, USA}
\affiliation{Center for Cosmology and Astroparticle Physics, The Ohio State University, 191 W. Woodruff Avenue, Columbus, OH 43210, USA}

\begin{abstract}
There is a longstanding discrepancy between the observed Galactic classical nova rate of $\sim 10$ yr$^{-1}$ and the predicted rate from Galactic models of $\sim 30$--50 yr$^{-1}$. One explanation for this discrepancy is that many novae are hidden by interstellar extinction, but the degree to which dust can obscure novae is poorly constrained. We use newly available all-sky three-dimensional dust maps to compare the brightness and spatial distribution of known novae to that predicted from relatively simple models in which novae trace Galactic stellar mass. We find that only half ($\sim 48$\%) of novae are expected to be easily detectable ($g \lesssim 15$) with current all-sky optical surveys such as the All-Sky Automated Survey for Supernovae (ASAS-SN). This fraction is much lower than previously estimated, showing that dust does substantially affect nova detection in the optical. By comparing complementary survey results from ASAS-SN, OGLE-IV, and the Palomar Gattini IR-survey in the context of our modeling, we find a tentative Galactic nova rate of $\sim 40$ yr$^{-1}$, though this could decrease to as low as $\sim 30$ yr$^{-1}$ depending on the assumed distribution of novae within the Galaxy. These preliminary estimates will be improved in future work through more sophisticated modeling of nova detection in ASAS-SN and other surveys.
\end{abstract}


\section{Introduction} \label{sec:intro}

A classical nova occurs in an interacting binary system with a white dwarf primary, referred to as cataclysmic variable (CV; see \citealt{warner_1995}). The white dwarf accretes material from the secondary, usually a low-mass main sequence star, until a critical pressure and temperature are reached, leading to a thermonuclear runaway at the bottom of the hydrogen-rich shell accreted by the white dwarf (see \citealt{be08} for a review). Expulsion of the accreted envelope occurs, causing the system to brighten significantly, by 5 to $>$19 mag \citep{kawash21}. Studies of classical novae in M31 have constrained the peak luminosities to range from M$_V$ $\approx$ $-4$ to $-10$ mag \citep{shafter_2017}. 

Historically, amateur astronomers have been the driving force in finding classical novae, with discoveries dating back thousands of years \citep{patterson13,shara17}. Novae began to be more systematically discovered in the mid 20th century, when on average $\sim$3 per year were visually discovered. When film photography became commonly used in the 1980s and 1990s, there were $\sim$4 discovered novae per year, and then $\sim$8 per year in the 2000s and 2010s when digital cameras became widely available. Then, in 2017, the All-Sky Automated Survey for SuperNovae (ASAS-SN) became the first survey to systematically observe the entire night sky with  nearly daily cadence down to $g \approx$ 18.5\,mag \citep{spg14}, significantly deeper than most amateur observations. Since 2017, there have been roughly 10 classical novae discovered per year on average.

The first prediction for the total frequency of Galactic nova eruptions was by \cite{lund35L}, who estimated there should be about 50 novae per year (see \citealt{di20} for a review of Galactic nova rate predictions). Estimates from the early 1990s predicted much lower rates  ranging from 11 to 20 per year, derived from observations of other galaxies \citep{cfw90,van91,della94}. More recent surveys of M31 have increased these predictions for the Milky Way rate to $34_{-12}^{+15}$ yr$^{-1}$ \citep{darnley06} and as high as $\sim50$ to $\sim70$ yr$^{-1}$ when accounting for incompleteness of faint and fast novae \citep{shafter_2017}. Recent work modeling novae in our Galaxy predict rates of $50_{-23}^{+31}$ yr$^{-1}$ from a sample of bright novae \citep{shafter_2017} and $43.7_{-8.7}^{+19.5}$ yr$^{-1}$ from a sample of novae detected in IR observations \citep{de21}. If these recent, higher estimates are correct, novae could be key contributors to various isotopes present in the Galaxy \citep{JOSE06} like  $^{26}$Al \citep{jose98,bennett13} and $^{7}$Li \citep{Starrfield78,hernanz96}, but there must be a reason the majority of novae go undetected.

So, what is the cause of the discrepancy between the predicted and the observed rate? One idea put forward is that the majority of observable classical nova events go undetected due to insufficient sky coverage of observations. However, the emergence of large sky surveys, including ASAS-SN, should solve this issue. The most common Galactic transient ASAS-SN discovers is a dwarf nova outburst, and \cite{kawash21} explored the possibility that some classical novae were being mistaken for dwarf novae. Though it is possible a small number of novae can be mis-classified, there are too few to significantly alter the discovery rate of classical novae. 

Another possibility is that interstellar dust obscures a majority of classical novae that erupt in the Galaxy. This prospect is supported by the recently discovered sample of highly reddened, and optically missed, novae discovered by the Palamor Gattini IR-survey (PGIR; \citealt{de20,de21}). These results suggest that dust could cause a substantial fraction of Galactic nova events to go undetected by optical observations, but how many remains an open question. 

\cite{shafter_2017} used an exponential disk to model extinction in the Galaxy, with 1 mag of extinction per kpc in the mid-plane in $V$-band. This predicts that over 90$\%$ of all Galactic novae should get brighter than $V =$ 18 mag, inconsistent with the results of \cite{de21}. The primary goal of this work is to investigate how utilizing a more robust Galactic extinction model changes these conclusions, and the availability of three-dimensional Galactic extinction maps now make it possible to explore this exact question.

Recent high angular resolution observations combined with stellar evolution models have resulted in several three dimensional extinction models of the Galaxy. \cite{ged15} used a combination of \textit{Gaia}, Pan-STARRS, and 2MASS observations to model the extinction north of $\delta = -30^\circ$, and \cite{marshall06} used 2MASS data to model extinction around the Galactic Center. Neither of these maps cover the entire sky, but \cite{bovy15} combined these maps along with analytic models from \cite{drimmel03} to build an all-sky three dimensional model of extinction in the Galaxy.

\cite{de21} used the \cite{green19} 3D dust map to build a model of Galactic novae in the PGIR field of view (decl. $> -28.9^\circ$), and compared their modeled novae to a sample of optically and IR discovered novae. The results suggest the optically discovered novae have much lower extinction than expected from their model and that the PGIR discovered sample was consistent with this modeled distribution of extinction. This bolsters the argument that dust is a significant factor in obscuring novae in the optical, but their analysis did not cover a large portion of the bulge and inner disk. Here, we extrapolate to the entire sky using the \cite{bovy15} all-sky dust map, to build upon these recent findings. 

The goal of this work is to model the distribution of novae and dust within the Galaxy, and explore how that combination affects how bright novae are when observed in the optical. In Section \ref{sec:model}, we discuss the implementation and assumptions of the model, including the resulting apparent magnitude distribution of novae, how many optically missed novae we should expect, and where to find them with IR observations or optical observations in redder bands. Then in Section \ref{rates}, we use our model along with observational constraints from various surveys to estimate the global frequency of nova eruptions. Lastly, in Section \ref{assume}, we explore how sensitive our results are to our model assumptions.

\section{Nova Model/Results}
\label{sec:model}
Here we discuss the components and assumptions that go into our Galactic nova model, the resulting Galactic apparent magnitude distribution of novae, and the spatial distribution of optically observable versus unobservable novae.

\subsection{Stellar Density Profile}
\label{sec:catun}
To analyze the effects of extinction on novae, we first must assume some distribution of novae within the Galaxy. For our primary model, we simply assume that the distribution of novae follows the distribution of stellar mass, but in Section \ref{sec:pop} we also consider additional models with a higher rate of nova production per unit mass in the bulge (as compared with the disk). The stellar mass distribution is inferred by implementing a bulge, a thin disk, and a thick disk component from the Contracted Halo version of the mass profile presented in \cite{catun20}. The distribution of mass is  calculated on a three-dimensional Cartesian grid with size (x,y,z) = (30,30,30) kpc and a resolution of 0.1 kpc. The total mass of each component is calculated and found to be consistent with the derived mass in Table 2 of \cite{catun20}. This was chosen for our preferred stellar density model due to the thorough explanation of the model parameters and derived values allowing us to confidently reproduce their work. In Section \ref{sec:mass}, we distribute novae based on a different stellar distribution model and this does slightly affect the resulting apparent magnitude distribution.

\subsection{Extinction Model}
\label{sec:dust}
To date, there is no single three-dimensional dust map that models Galactic extinction across the entire sky. However, the maps of \cite{ged15}, \cite{marshall06}, and \cite{drimmel03} were combined and made publicly available at \url{http://github.com/jobovy/mwdust} to provide a stitched together map over the entire sky \citep{bovy15}. This model, hereafter referred to as \texttt{mwdust}, can use several different map combinations, and here we use the \texttt{combined19} version. This uses the updated map of the sky north of declination $\delta = -30^\circ$ from \cite{green19}, the \cite{marshall06} maps covering the sky around the Galactic center $-100^\circ \leq l \leq 100^\circ$ and $-10^\circ \leq b \leq 10^\circ$, and the \cite{drimmel03} map for the rest of the sky not covered by the first two. The \cite{marshall06} map takes precedence over the \cite{green19} map where they overlap, because the latter was found to underestimate the amount of extinction at low latitude. A detailed explanation of the model can be found in the Appendix of \cite{bovy15}.

\subsection{Positions and Distances}
\label{sec:pos}
\begin{figure}
\begin{center}
 \includegraphics[width=0.48\textwidth]{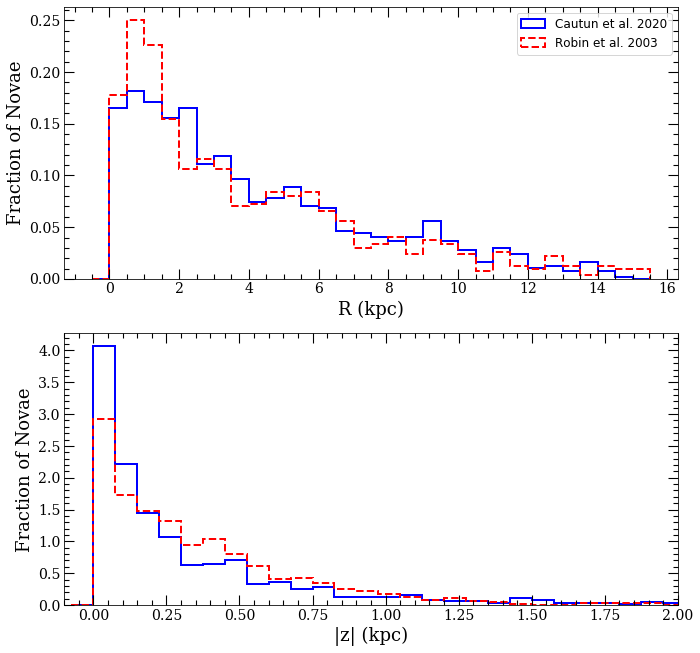}
\caption{Distributions of Galactic positions in cylindrical radii (top) and height above the disk (bottom) for 1000 randomly sample novae distributed by randomly sampling from the \cite{catun20} shown as the blue solid line and from the \cite{robin03} model shown as the  red dashed line.}
\label{fig:rz_dist}
\end{center}
\end{figure}

\begin{figure}
\begin{center}
 \includegraphics[width=0.48\textwidth]{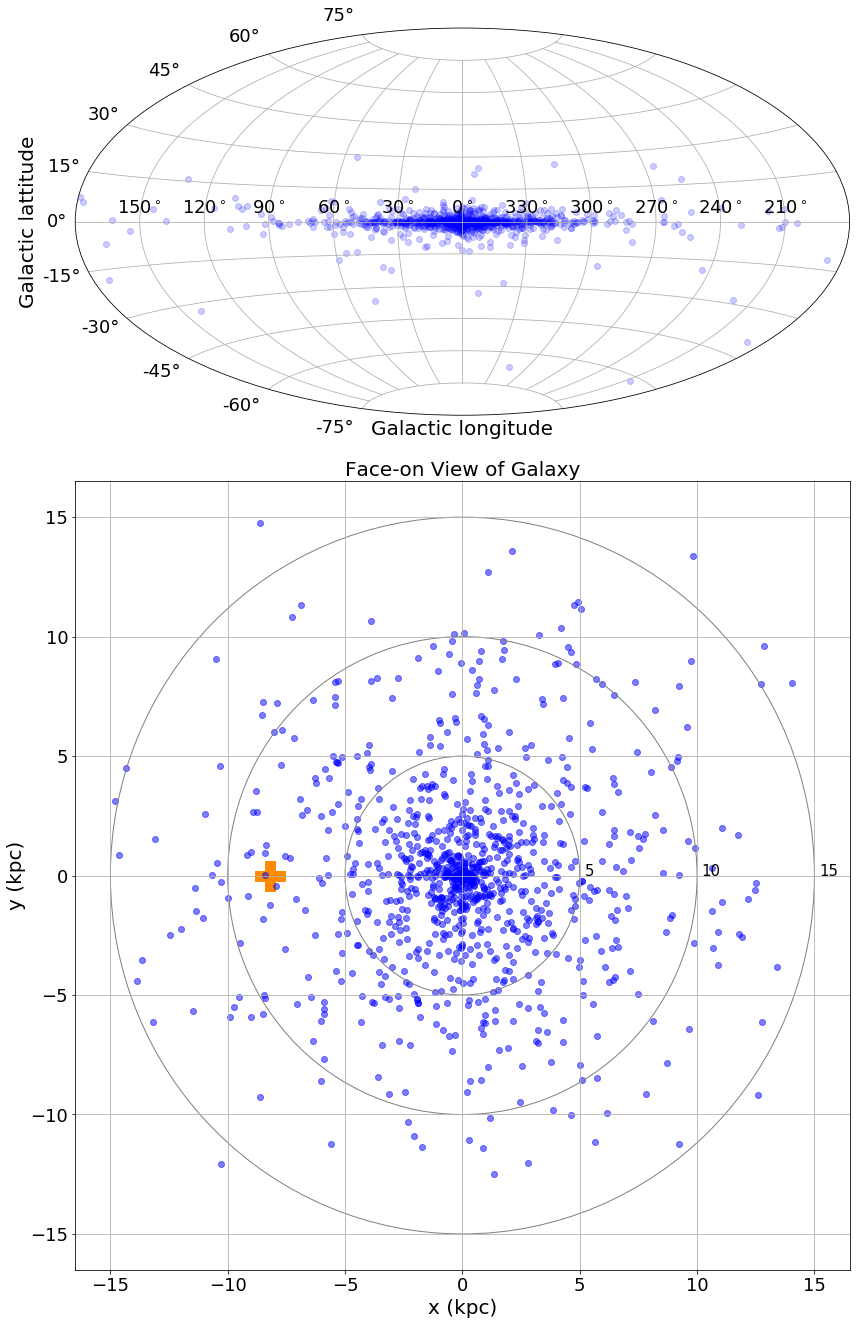}
\caption{Top: positions of N = 1000 simulated novae in Galactic coordinates, distributed by randomly sampling the \cite{catun20} stellar density model.
Bottom: same as top but in an external face-on view of our Galaxy. Novae
are plotted in blue, and the position of the Sun is as an orange cross.}
\label{fig:skymap}
\end{center}
\end{figure}

We ran Monte Carlo simulations of 1000 Galactic novae by probabilistically distributing them following the stellar mass of the Galaxy from \cite{catun20}. The distribution of novae in Galactic cylindrical coordinates for the primary \cite{catun20} model and a secondary stellar density model are shown in Figure \ref{fig:rz_dist}. In Figure~\ref{fig:skymap}, these positions are transformed to sky coordinates at the reference frame of the Sun R$_\odot = 8.122$ kpc \citep{gravity_collab,catun20} in the top panel, and a face-on view of the Galaxy is shown in the bottom panel. As expected, novae hug the disk plane, and there is a large increase in density toward the Galactic center. 

\begin{figure}
\begin{center}
 \includegraphics[width=0.48\textwidth]{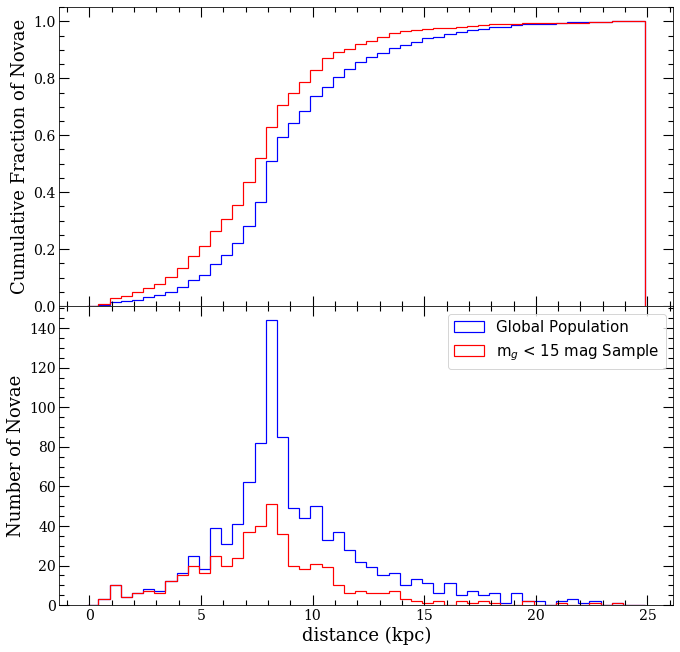}
\caption{Distribution of distances from the Sun for N = 1000 simulated novae. The top panel shows a cumulative distribution, and the bottom panel is a normalized histogram.}
\label{fig:distances}
\end{center}
\end{figure}

The distances to Galactic novae are often hard to constrain even with improved parallax techniques from \textit{Gaia} \citep{Shaefer18}. Figure \ref{fig:distances} shows the expected distribution of distances based on our model. We expect the median distance to a nova to be 8.5 kpc, 68$\%$ of novae to have distances between 6.0 and 12.2 kpc, and 95$\%$ of novae are expected to be within 15.8 kpc of the Sun. Also shown in Figure \ref{fig:distances} is the distribution of distances from a magnitude limited sample of novae that we expect to more closely resemble the sample of optically discovered novae. The median distance of this distribution is 7.9 kpc, very similar to the global population, and we expect 95$\%$ of this magnitude limited sample to be within 14 kpc. Overall, the distribution of distances for a magnitude limited sample is not significantly different than the global population, suggesting that distance alone is not the determining factor for missed novae.

\subsection{Brightness Distribution}
\label{sec:bright}

The peak absolute magnitude of each nova is randomly sampled from a normal distribution with a mean and standard deviation of $\mu = -7.2$ mag and $\sigma = 0.8$ mag, respectively \citep{shafter_2017}. This distribution was derived for M31 novae in the $V$-band, and we assume there is no magnitude difference between this and Milky Way novae at $g$-band peak \citep{vandenbergh87,miro88,hachisu14}. In deriving Galactic nova rates both \cite{shafter_2017} and \cite{de21} explore altering the luminosity function for the bulge and disk novae together and separately. We only assume the above luminosity function for all of the novae in our model, as \cite{de21} found it is not a significant factor for the results However, we plan to investigate the effects of the luminosity function in Kawash et al.\ 2022 (in preparation). After a Galactic position was randomly assigned to the nova, the accompanying extinction for that line of sight and distance was estimated using the \texttt{mwdust} package. These values were combined with the distance and randomly assigned absolute magnitude
to estimate the peak apparent magnitude for each nova. 

\begin{figure}
\begin{center}
 \includegraphics[width=0.48\textwidth]{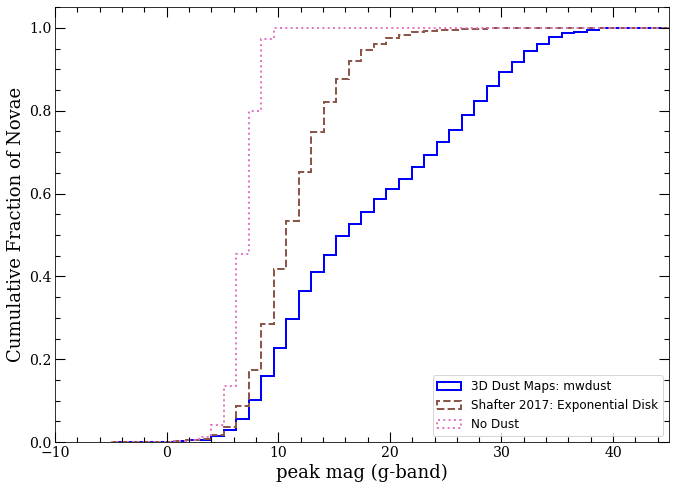}
\caption{Cumulative distributions of peak apparent magnitudes of N = 1000 simulated novae. Three models are shown: one that includes no dust (pink dotted line), one with an exponential dust disk as implemented in \cite{shafter_2017} (brown dashed line), and the $\texttt{mwdust}$ model (blue solid line).
The different models yield significantly different distributions, highlighting the importance of accurately modeling dust to estimate the nova rate.}
\label{fig:mag_dist}
\end{center}
\end{figure}

The cumulative distribution of peak apparent magnitudes of Galactic novae is shown in blue in Figure \ref{fig:mag_dist}, and compared to a distribution excluding dust and a distribution implementing the disk extinction model of \cite{shafter_2017}. It is clear that the exponential disk utilized in \cite{shafter_2017} vastly underestimates the effects of dust relative to the estimates from three-dimensional dust maps. Specifically, modeling extinction from three-dimensional dust maps predicts that only  $48\%$ of novae in the Galaxy will get brighter than $\textit{g} = 15$ mag, while \cite{shafter_2017} predicted $82\%$ of novae will be brighter than $\textit{g} = 15$ mag. This could explain why ASAS-SN observations have not resulted in a significant increase in the nova discovery rate, and it is consistent with the scenario that a large fraction of nova eruptions are too faint to be detected in blue optical bands but are detectable in the IR, like the recently discovered PGIR sample \citep{de21}. Also, this likely means the deeper observations of the Vera C. Rubin Observatory Legacy Survey of Space and Time (LSST; \citealt{lsst}) will discover many more Galactic novae than previously thought. 

The accuracy of this distribution relies heavily on the ability of the 3D dust maps to estimate high extinctions at low latitudes out to large distances. A majority of the novae (94$\%$) are in the area of the sky that the \cite{marshall06} map covers, and for regions at high column densities, this map only has information out to $\sim$7 kiloparsecs. 5$\%$ of novae fall within the \cite{green19} region, and this model only extends to a few kiloparsecs. The remaining few novae lie in the \cite{drimmel03} region, where the analytic model extends out to a galactocentric radius of $R = 15$ kpc, or the entire size of the grid. So, a large fraction of the novae in our model could have underestimated extinction values, but we suspect this is only for the severely extinguished novae, and thus they are already unobservable. Therefore, we do not expect that our prediction that only $48\%$ of novae in the Galaxy get brighter than $\textit{g} = 15$ mag would change if the extinction model was complete for the entire Galaxy, although it could change predictions for IR surveys and observations carried out in redder bands.

\subsection{Reddened Novae}
\label{sec:red}

\begin{figure}
\begin{center}
 \includegraphics[width=0.48\textwidth]{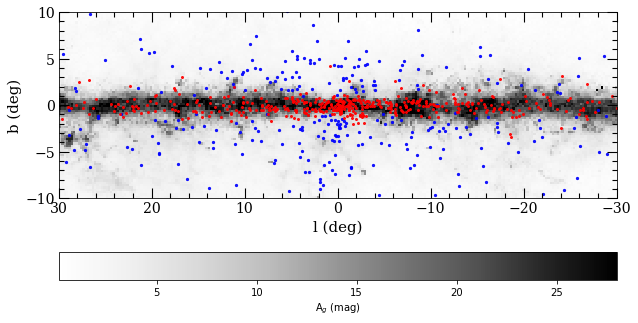}
\caption{Galactic coordinate positions of N = 1000 simulated novae around the Galactic center. The blue points indicate novae that reach a peak brightness of $g = 15$ mag or brighter, and would likely be discovered by optical observations. The red points indicate novae that never reach g = 15 mag and have a much lower chance to be discovered by optical observations. The amount of extinction integrated out to 15 kpc from the $\texttt{mwdust}$ model is shown as a grey scale color map. The resolution of the map is 0.25 degrees with a maximum extinction value of 28 mag for visualization purposes.}
\label{fig:map_model}
\end{center}
\end{figure}

Figure \ref{fig:map_model} shows the positions of our modeled novae in Galactic coordinates around the Galactic center, distinguished by whether the peak apparent magnitude reached $g = 15$ mag. As expected, almost all of the heavily obscured, and therefore faint, novae lie within a couple degrees of the Galactic plane. This implies that optical observations will struggle to discover novae in regions within a couple degrees of the plane and especially towards the Galactic center. 

To explore how our model predictions compare to the known sample of optically discovered novae, we have compiled a list of known novae by combining the sources from the CBAT list of novae in the Milky Way\footnote{\url{http://www.cbat.eps.harvard.edu/nova_list.html}} and Koji's List of Recent Galactic Novae\footnote{\url{https://asd.gsfc.nasa.gov/Koji.Mukai/novae/novae.html}}. The CBAT list consists of objects from 1612 to 2010, and we only include objects with eruptions after 1900. A literature search was then preformed on the entire list to investigate if any contaminating sources were present and how many objects have been spectroscopically confirmed as novae. We found that 10 objects from the CBAT list are not classical novae and have removed them from our list. 351 of the objects have spectroscopic or photometric observations suggesting they are indeed classical novae, but we find no information about the remaining 47 objects. We assume that these objects are classical novae, but the possibility of contamination still remains. 

\begin{figure*}
\begin{center}
 \includegraphics[width=1.0\textwidth]{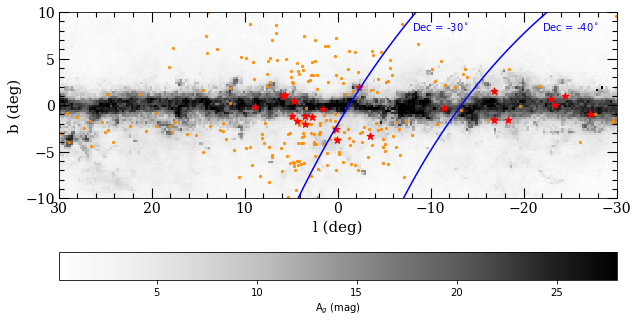}
\caption{Positions of known, optically discovered novae in Galactic coordinates around the Galactic center as orange dots. Nova candidates reported by VVV and OGLE observations are shown as red stars. VVV and OGLE should be better suited for finding reddened novae in the plane than bluer optical observations,  but OGLE has a lower cadence in these highest extinction regions. The dust map is the same as Figure \ref{fig:map_model}. Lines of constant declination are shown in blue to highlight the lack of optically discovered novae at the most southern declinations.}
\label{fig:map_sample}
\end{center}
\end{figure*}

Figure \ref{fig:map_sample} shows the positions of optically discovered novae from our list in Galactic coordinates towards the Galactic center. Consistent with predictions from our model, there appears to be a significant lack of novae near the Galactic plane where most of the obscuring dust resides ($|b| \lesssim 2^{\circ}$). However, there also appears to be a bias against discovery of novae at lower declination. This is likely due to a historic lack of observations in the Southern Hemisphere, although this issue should have recently been addressed by ASAS-SN's Southern Hemisphere facilities and an increase in amateur observers in Australia and Brazil (e.g., the Brazilian Transient Search, BraTS). 

Also shown in Figure \ref{fig:map_sample} are nova candidates discovered by The Vista Variables in the Via Lactea (VVV; \citealt{vvv10}) and Optical Gravitational Lensing Experiment (OGLE; \citealt{ogle4}). The VVV deep near-IR observations and the OGLE $I$-band observations of the Galactic bulge and nearby disk are better suited than most optical observations to discover novae in dustier fields at low Galactic latitudes, but the cadence of OGLE observations is much lower in these high extinction regions at low lattitude \cite{udalski15}. The 20 candidates from VVV \citep{vvv_atel12a,vvv_atel13a,vvv_atel13b,vvv_atel13c,vvv_atel14d,vvv_atel15a,vvv_atel15b,vvv_atel16a,vvv_atel16b,vvv_atel16c,vvv_atel17a} and 19 from OGLE \citep{ogle_atel12a,ogle_atel14a,ogle_atel14b,ogle_atel14c,ogle_atel16a} were either discovered in the data after eruption or not followed up spectroscopically. Many are likely classical nova eruptions, though the sample could be contaminated by a few dwarf novae and young stellar objects.  As seen in Figure \ref{fig:map_sample}, it does appear to be the case that there are more VVV and OGLE nova candidates closer to the plane, but there are still regions where few to no novae or nova candidates have been discovered. 

For example, there has never been a nova or nova candidate discovered in the 20 deg$^2$ patch of sky with a Galactic longitude and latitude of $-10^\circ < l  < 0^\circ$ and $-1^\circ < b < 1^\circ$, respectively. Our model predicts that $\sim10\%$ of Galactic novae should be in this region, and there is A$_g$ = 23 mag of extinction on average according to the \cite{marshall06} dust map. This is a region that OGLE observed less frequently than other bulge fields \cite{udalski15}, but ASAS-SN has observed this region at a high cadence.

\begin{figure}
\begin{center}
 \includegraphics[width=0.48\textwidth]{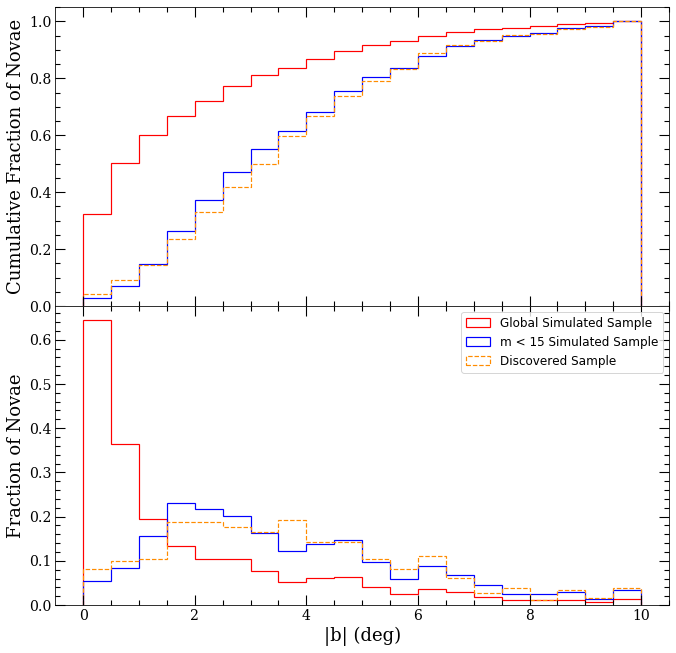}
\caption{Normalized cumulative distribution (top) and normalized histogram (bottom) of novae as a function of Galactic latitude. All simulated novae are plotted in red, while simulated novae that reach a brightness of $g = 15$ mag are shown in blue. Optically discovered novae are plotted as an orange dashed line. The discovered sample more closely resembles the bright m < 15 mag model, suggesting a bias against discovering novae in regions of heavy extinction and a severe historic lack of novae discovered at low Galactic latitude.}
\label{fig:lat}
\end{center}
\end{figure}

The absence of novae at low Galactic latitude is perhaps shown more clearly in Figure \ref{fig:lat}, which compares the  Galactic latitude distribution of simulated novae with known optically discovered novae. Our model predicts that $\sim 65\%$ of novae erupt within $\left | b \right |<2^\circ$; however, only $\sim20\%$ of the optically discovered sample resides within this region. Also shown in Figure \ref{fig:lat} is the distribution from a magnitude-limited sample of modeled novae that reach an apparent magnitude of $g = 15$ mag. This distribution peaks at a Galactic latitude of 2 degrees, similar to the observed distribution, consistent with a historic magnitude-limited sample with dust as the determining factor. 

\section{Global Nova Rate Estimates}
\label{rates}
Next, we discuss how our nova model, described in Section \ref{sec:model}, can be used to explore what fraction of simulated novae would be observable for various surveys and what that implies for the global Galactic nova rate.

The all-sky and nearly one day cadence observations of ASAS-SN provide an unprecedented opportunity to better constrain the Galactic nova rate. Even though the limiting magnitude of ASAS-SN is as deep as $g \sim 18$ mag, nova searches have some unique challenges and we believe the current infrastructure of the transient candidate pipeline is best suited to discover novae brighter than g $\sim$ 15 mag for various reasons. First, ASAS-SN usually flags portions of the sky known to contain variable stars, and avoids searching for transients in these regions---but classical novae often have variable hosts. To address this issue, 
a special ``ASAS-SN Nova Alert" email is generated for essentially all transients $g \lesssim$ 15 mag, regardless of previous variability, and is immediately sent to alert several of the co-authors upon detection. Second, confusion from neighboring sources---many of which are variable---is a much larger issue in the Galactic plane, and essentially translates to a shallower detection limit. Additionally, the candidate pipeline dedicated to discovering novae has a cutoff at this threshold. And finally, the number of contaminants greatly increases closer to the detection limit, so candidates fainter than $g = 15$ mag are not always checked. ASAS-SN can, and does, find Galactic transients fainter than this threshold, but the detection efficiency likely falls off quickly at $g > 15$ mag. This detection threshold is bright, especially when compared to the number of fainter CV candidates and extragalactic supernova candidates discovered by the survey\footnote{http://www.astronomy.ohio-state.edu/asassn/transients.html}, and the detection efficiency of crowded fields at deeper thresholds will be explored in future work.

Currently, the detection efficiency of Galactic transients brighter than $g \approx 15$ mag is not known for ASAS-SN. For extragalactic SNe, \cite{holoien19} found that ASAS-SN is essentially complete down to $m = 16.2$ mag, but we do not expect the completeness to be as deep in the plane where almost all novae reside. To estimate this value, a fake transient recovery analysis performed on ASAS-SN data is needed, but it is beyond the scope of this work. Even with perfect recovery in observable fields, an optical transient survey is limited to detection rates $\lesssim80\%$ due to Solar conjunction \citep{mroz15}, and preliminary estimates of detecting fainter, but longer lived, Type~Ia supernovae in ASAS-SN suggest detection capabilities between $70\%$ and $80\%$ (Desai et al.\ 2021, in prep). A fake transient recovery analysis was performed on PGIR data in \cite{de21}, where it was estimated that $36\%$ of all Galactic novae that reach $J = 14$ mag in their field of view could be detected ($\delta > -28.9^\circ$ at a cadence of $\approx$ 2 nights). This lower detection capability is largely due to crowding/blending from an 8" pixel scale, contamination from nearby bright stars, and the Galactic center being unobservable for a large fraction of the year from PGIR's Mt. Palomar location. We expect the first two issues to be present in ASAS-SN data since the two surveys have the same pixel scale and a majority of novae should be found within a couple degrees of the crowded plane, but the last issue is not as severe because ASAS-SN has facilities in both the Northern and Southern Hemisphere. Taking all of the above information, we estimate ASAS-SN can detect $60\%$ of Galactic novae that reach $g = 15$ mag. To be clear, this estimated detection efficiency is only applicable to Galactic novae, and the detection efficiency should be higher for extragalactic transients in less crowded fields off of the plane. 

Between 2018--2020, there were 31 known Galactic novae  that peaked brighter than $g = 15$ mag. A majority of these were clearly detected and flagged as transients in ASAS-SN data, but there are at least a few examples of novae that were not detected or flagged as nova candidates. V3731 Oph \citep{v3731_atel} was detected as a transient candidate in the ASAS-SN pipeline but was confused with a coincident variable within a pixel of the nova. V6567 Sgr \citep{V6567_atel} was detected on the rise in ASAS-SN data and initially reported as a CV candidate. And though V1709 Sco \citep{v1709_atel} was detected as bright as $V = 12.7$ mag by other observers, it was never detected brighter than $g = 15$ mag in ASAS-SN data likely due to facilities being shutdown for a large portion of 2020 due to the pandemic. It is likely that more observable nova events were missed by all transient surveys and observers and even more due to solar conjunction, so this is consistent with our $60\%$ detection efficiency estimate for this time period. This estimation is very crude, and it will be one of the major goals of Kawash et al.\ (2022 in preparation) to better understand and constrain it. 

We use our model to distribute N = 1000 novae in a mock galaxy, estimate what fraction would be detectable by ASAS-SN by assuming detections of 60$ \pm 6\%$ that reach g = 15 mag, and extrapolate to a global rate from an annual discovery rate of R $= 10.3 \pm  1.9$.  This analysis is carried out for 1000 iterations, each time sampling a normal distribution with a mean and standard deviation equal to the estimated value and uncertainty, respectively, for each parameter in order to evaluate  the most likely Galactic nova rate based on ASAS-SN observations.

Since this is the first nova rate estimate from ASAS-SN, and because the detection efficiency is not well constrained, we compare our results to those derived from other transient surveys. First, it was estimated that between 2010--2013 OGLE-IV observations discovered up to 80$\%$ of novae brighter than $I = 17$ mag in the most frequently visited fields  in their field of view  $(- 10^\circ < l <  10^\circ$ and $- 7^\circ < b <  5^\circ$, with a cadence varying from 20 minutes to a few days; \citealt{mroz15}). There was a discovery rate of R $= 4.8 \pm 1.1$ yr$^{-1}$ over this time period. We carry out the same analysis as we did for ASAS-SN to derive a Galactic nova rate from OGLE-IV observations using a detection efficiency as a function of sky position estimated from Figure 9 of \cite{mroz15}, but we ignore novae directly in the plane ($\left | b \right | < 1^\circ$) and fields with a detection efficiency less than $25\%$ to account for the different extinction model used in our analysis. We believe this is a safe assumption as only one nova from Table 1 of \cite{mroz15} lies within this ignored region.  Our model estimates that OGLE-IV observations detected roughly $34\%$ of novae within $-10^\circ < l <  10^\circ$ out to 10 kpc, consistent with the $36\%$ detection rate of bulge nova estimated in \cite{mroz15}. OGLE-IV observations have a much lower cadence in the highest extinction regions in the plane \cite{udalski15}, so a large fraction of novae are likely undetected despite the better pixel scale and redder filter compared to ASAS-SN. However, roughly $40-50\%$ of these novae are too highly extinguished to be detectable even with improved monitoring of the field.

Another survey with a published rate of nova discovery is the PGIR survey \citep{de21}. Over the first 17 months of observations, they discovered 7.8 novae per year and estimated they could detect $36\%$ of all novae brighter than $J \sim$ 14 mag in their field of view ($\delta > -28.9^\circ$ at a cadence of $\approx$ 2 nights). We run our analysis on a PGIR detection rate of $r = 7.8 \pm 2.3$ yr$^{-1}$ and detection efficiency of $\epsilon = 0.36 \pm 0.036$ in the field of view.

Lastly, \cite{shafter_2017} derived a Galactic nova rate with a bright, nearby sample of novae. Between 1900 and 2020, there have been only 7 novae that reached an apparent magnitude of $m = 2$. We carry out our analysis assuming that $90\%$ of all novae that reached this brightness were discovered. 

 \begin{figure}
\begin{center}
 \includegraphics[width=0.48\textwidth]{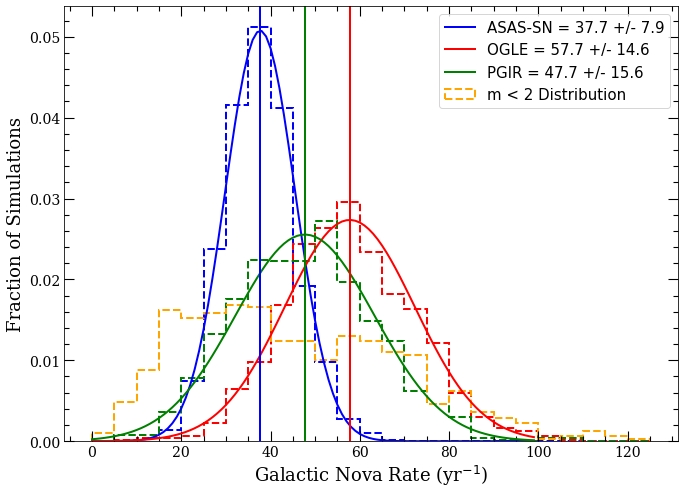}
\caption{Distribution of the Galactic nova rate from 1000 iterations of our model based on observational constraints from ASAS-SN (blue), OGLE (red), PGIR (green), and historic bright novae (orange) shown as dashed histograms. The results from ASAS-SN, OGLE, and PGIR observations are well fit by a normal distribution, and this is shown for each respective survey along with the mean value.}
\label{fig:catun1}
\end{center}
\end{figure}

The distributions of nova rates derived from these various observational constraints are shown in Figure~\ref{fig:catun1}.  
All of these distributions except for the $m < 2$ mag constraint are well fit by a normal distribution, and for those we derive the mean and standard deviation (listed in Table~\ref{table:rates}). The ASAS-SN, OGLE, and PGIR derived rates are all consistent at the 1$\sigma$ level. The OGLE-IV rate ($57.7 \pm 14.6$ yr$^{-1}$)  and PGIR rate ($48.3 \pm 15.8$ yr$^{-1}$) have a higher variance because of their limited fields of view. OGLE-IV has lower cadence at the lowest latitude (where most novae are in our model), and PGIR is unable to observe the Galactic center for a large portion of the year. The bright nova constraint ($m < 2$ mag) results in a distribution with high variance due to a low observed rate (7 novae over 120 years).
The ASAS-SN derived distribution ($39.7 \pm 8.4$ yr$^{-1}$) results in the distribution with the lowest variance, but this distribution is derived without analyzing  the detection efficiency of novae robustly. Once this value is better constrained, and more observations are accumulated, ASAS-SN could provide the best constraint on the Galactic nova rate.

\begin{deluxetable*}{cccc|cccc}
\tablecolumns{8}
\tablewidth{20pt}
\tablecaption{Galactic Nova Rates For Various Parameters}
\label{table:rates}
\tablehead{
\colhead{} & 
\colhead{} & 
\colhead{Parameters} &
\colhead{} & 
\colhead{} & 
\colhead{Implied Galactic Rate (yr$^{-1}$)} &
\colhead{} & 
\colhead{} \\
\colhead{$\theta$} & 
\colhead{Dust Model} & 
\colhead{Mass Model} &
\colhead{$N_d/N_b$} & 
\colhead{ASAS-SN} & 
\colhead{OGLE} &
\colhead{PGIR} & 
\colhead{$\Bar{\chi}$} 
}
\startdata
1.0 & mwdust & Cautun & 5.0 & $38 \pm 8$ & $58 \pm 14$ & $48 \pm 16$ & 43  \\
1.0 & mwdust & Besancon & 1.70 & $33 \pm 7$ & $38 \pm 9$ & $47 \pm 16$ & 36  \\ 
1.0 & Exponential & Cautun & 5.0 & $26 \pm 5$ & $57 \pm 14$ & $45 \pm 15$ & 32 \\ 
0.4 & mwdust & Cautun & 1.67 & $38 \pm 8$ & $44 \pm 11$ & $48 \pm 15$ & 42 \\
0.4 & mwdust & Besancon & 0.67 & $31 \pm 7$ & $29 \pm 7$ & $47 \pm 15$ & 33  
\enddata
\tablecomments{Model parameters and derived Galactic nova rates based on ASAS-SN, OGLE, and PGIR detections of novae. $\theta$ is the ratio of disk to bulge novae per unit mass and $N_d/N_b$ is the resulting disk to bulge ratio of novae for a given mass model. $\Bar{\chi}$ is the weighted average of the Galactic nova rate from the three surveys.}
\end{deluxetable*}

The observational constraints from these various surveys are almost completely independent. The OGLE rate is derived from OGLE-IV observations, occurring between 2010--2013. The ASAS-SN discoveries occur between 2018--2020, overlapping with the PGIR discoveries from July 2019 $-$ November 2020. There have been no m = 2 mag novae that have erupted since any of these surveys started observing. A simple weighted average of ASAS-SN, OGLE, and PGIR derived rates results in a Galactic nova rate of $R \approx 43$ yr$^{-1}$.

\section{How Sensitive Are The Results to our Assumptions?}
\label{assume}

Here we explore how our results change as we vary certain assumptions in our model. In Section~\ref{sec:shaft_dust}, we change the extinction model from the 3D dust maps to the simple exponential used in \cite{shafter_2017}. Then  in Section~\ref{sec:mass}, we see how our results depend on the mass model of the Galaxy. Finally, we briefly explore how assuming different populations of bulge and disk novae affects our results in Section~\ref{sec:pop}. The derived rates from various sets of parameters are shown in Table~\ref{table:rates}.

\subsection{Extinction Models}
\label{sec:shaft_dust}
The dust maps of \cite{green19}, \cite{marshall06}, and \cite{drimmel03}---stitched together by \cite{bovy15}---form the best all-sky three dimensional dust map to date. It is almost certainly superior to simply assuming a disk of dust (i.e., as in \citealt{shafter_2017}), but it is only able to model extinction out to a few to $\sim 10$ kpc depending on the direction. For this reason, we investigate how the rates change if we implement the exponential disk model of extinction used in \cite{shafter_2017}. The results are shown in the top right panel of Figure \ref{fig:rate_comp}.

As expected, the rate estimated from the $g$-band observations of ASAS-SN is extremely sensitive to the dust model used. The exponential model of extinction underestimates the amount of dust in the plane relative to the $\texttt{mwdust}$ model, therefore yielding a higher detection fraction and ultimately a much lower rate. The OGLE derived rate is not sensitive to the dust model since we assume they do not detect novae in the fields with the highest extinction, and the IR observations of PGIR are also not sensitive to the dust model. The rates derived from the three surveys are no longer consistent when using this dust model, and it is clear that an under- or over-estimation of Galactic extinction will cause a significant error in the derived nova rate from ASAS-SN observations. This conclusion is consistent with predictions for observing the next Galactic supernova. \cite{adams13} found that different dust models yield different likelihoods of observing a Galactic supernova in the optical but was less important for near-IR observations.

\subsection{Mass Model}
\label{sec:mass}

The \cite{catun20} stellar density model is just one of many widely used Galactic mass models, so we explored if our results change if we use another model. Another commonly used model is the Besan\c con stellar density model first outlined in \cite{robin03}. We implement a version of this model that contains a two component bulge (discussed in \citealt{sbi17}), a thin disk, a thick disk, and a halo component; their total masses and normalization values are shown in Table \ref{table:comp_mass}. The form and parameters of each component can be found in Appendix \ref{sec:besancon}.

\begin{figure*}
\begin{center}
 \includegraphics[width=1.0\textwidth]{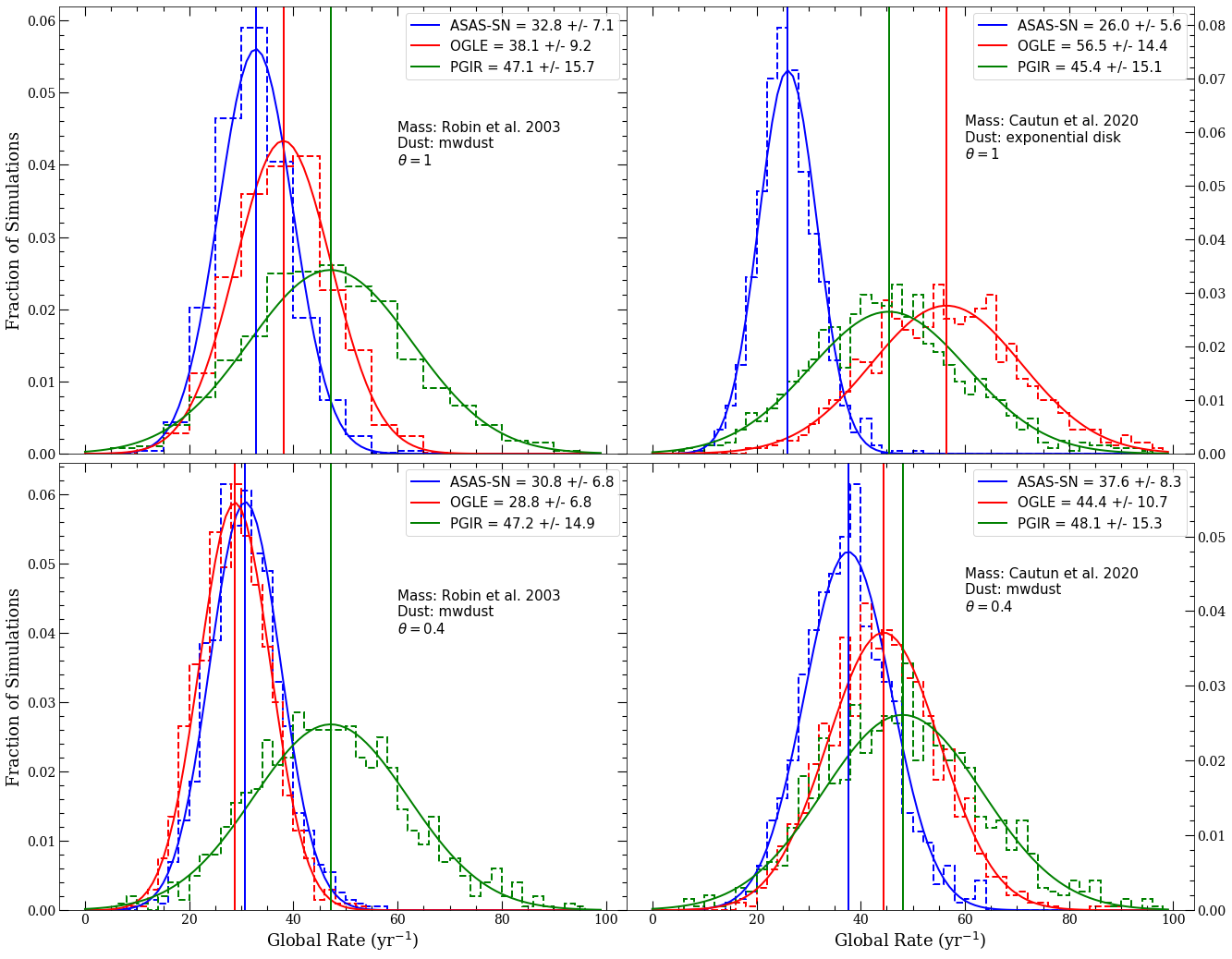}
\caption{Same as Figure \ref{fig:catun1} for various combinations of stellar density models, extinction models, and ratio of disk to bulge novae per unit mass ($\theta$). Top Left: Galactic nova rate distributions resulting from using the Besan\c con mass model. Top Right: Galactic nova rate distributions resulting from changing the $\texttt{mwdust}$ model to an exponential disk; this results in rate estimates inconsistent at the one-sgima level. Bottom Left: Galactic nova rate distributions from an elevated rate of nova production in the bulge using the \cite{robin03} mass model. Bottom Right: same as bottom left but for the \cite{catun20} model} Overall, the ASAS-SN and OGLE derived rates are sensitive to the model assumptions but the PGIR rate is not.
\label{fig:rate_comp}
\end{center}
\end{figure*}

The Besan\c con implementation of the stellar has a more massive, bar-like bulge component compared to the \cite{catun20} model. As seen in Figure \ref{fig:rz_dist}, the \cite{robin03} model places more novae at shorter Galactic radii but fewer at shorter Galactic height from the plane. This results in more bulge novae but slightly fewer novae in the highest extinguished regions, and, as seen in Figure \ref{fig:rate_comp}, this predicts a slightly lower rate from ASAS-SN observations and a significantly lower rate from OGLE observations. The PGIR rate does not appear to be sensitive to the stellar distribution model. Overall, using the \cite{robin03} model results in a lower prediction of the Galactic nova rate. but it is still consistent at the 1$\sigma$ level with using the \cite{catun20} model.

\subsection{Differing Bulge and Disk Populations}
\label{sec:pop}
It has been posited that novae that erupt in the bulge have different properties than those that erupt in the disk because of different progenitor populations hailing from differing star formation histories \citep{di20}. \cite{darnley06} found that the favoured model of novae in M31 supported separate disc and bulge populations  that erupted at different rates per unit $\textit{r}$-band flux. They found that, per unit  $\textit{r}$-band flux, the ratio of disk novae to bulge novae was 0.18. \cite{shafter01} also studied the spatial distribution of novae in M31, and estimated this ratio to be 0.4. In a similar fashion, we define $\theta$ as the ratio of disk novae to bulge novae per unit mass in our model. So far, we have assumed one population of novae that traced the overall stellar mass of the Galaxy ($\theta = 1$), resulting in a ratio of disk-to-bulge novae of $N_{\rm{disk}} / N_{\rm{bulge}} \approx 5$ for the \cite{catun20} mass model and $N_{\rm{disk}} / N_{\rm{bulge}} \approx 1.7$ for the \cite{robin03} mass model. 

Does the Milky Way Galaxy produce more novae per unit mass in the bulge than in the disk? Because distances are often hard to constrain, this is not an easy question to answer, but to first order, the higher the nova rate in the bulge, the more novae we should expect to find near $l = 0^\circ$. From our model where novae simply trace the stellar mass of the Galaxy ($\theta = 1$), we expect $40\%$ of bright ($g < 15$ mag) novae to be located within $\left | l \right | \leq 10^\circ$. Of all the known Galactic novae from our list, $45\%$ have erupted within $\left | l \right | \leq 10^\circ$. This could support the idea of bulge enhancement, or the bulge producing more novae per unit mass relative to the disk, especially since our discovered sample of novae is likely biased towards nearby disk novae. 

For this reason, we explore how our results differ from an elevated bulge rate of $\theta = 0.4$. This model predicts that $49\%$ of bright ($g < 15$ mag) novae are within $\left | l \right | \leq 10^\circ$ based on the \cite{catun20} mass model and  $67\%$ for the \cite{robin03} model. The value of $\theta$ is difficult to constrain due to the unknown number of foreground disk novae, but the number of known novae around the Galactic Center suggests it is larger than $\theta = 0.4$. The nova rate results for this elevated bulge distribution of novae are shown in the bottom panels of Figure \ref{fig:rate_comp} for both the \cite{catun20} and \cite{robin03} mass models.

If the production of novae in the bulge is elevated relative to the disk per unit mass, the global rate based on OGLE-IV observations decreases. This is expected, as OGLE observations are heavily biased to finding bulge novae. ASAS-SN and PGIR observations are less sensitive to the ratio of disk to bulge novae, as the field of view of these surveys are less biased towards the bulge or disk. 

\section{Conclusions}
We have used an all-sky three dimensional dust map to explore the effects of extinction on the discovered nova rate. This model predicts that roughly half of nova eruptions will be too faint for current ASAS-SN discovery abilities,  much higher than previous estimates that used a simpler dust model, and likely explaining much of the discrepancy between observed and predicted rates. Many of the highly extinguished, reddened novae lie within two degrees of the plane. Our model predicts that $\sim 65 \%$ of all novae should erupt within two degrees of the plane, compared to only $\sim 20 \%$ of the discovered sample being found in this region. This further highlights the necessity of optical surveys observing in redder bands and IR transient surveys like PGIR to detect these highly reddened novae in the plane, although these fields have lower recovery rates for surveys with large pixel scales. 

For the first time, we have estimated a Galactic nova rate based on an all-sky survey with nightly cadence.  ASAS-SN observations between 2018 and 2020 suggest a Galactic nova rate of $38 \pm 8\ \rm{yr}^{-1}$. This derived rate relies heavily on the detection efficiency (assumed to be $60\%$ in this work for novae brighter than $\textit{g} < 15$ mag), a value that will need to be better understood in the future. However, the derived rate from ASAS-SN is consistent with rates derived from OGLE observations (R = $58 \pm 14\ \rm{yr}^{-1}$) and PGIR observations (R = $48 \pm 16\ \rm{yr}^{-1}$). Our results are consistent with the recent higher rates derived in \cite{de21} and \cite{shafter_2017} but could support lower rates depending on how novae are distributed in the Galaxy.

The derived rate from ASAS-SN's blue $g$-band filter is sensitive to the extinction model implemented, so a precise nova rate estimation will rely on how accurately dust is modeled close to the plane. Similarly, the OGLE rate is sensitive to the level of bulge enhancement and along with the ASAS-SN rate is sensitive to the model used to place novae within the mock Galaxy. The PGIR rate does not appear to be sensitive to altering any of the assumptions of our model. For any combination of stellar density model, extinction model, and level of bulge enhancement, the observations of ASAS-SN, OGLE, and PGIR suggest a Galactic nova rate of $\sim30$ to $\sim40$ per year.

Overall, this work makes significant progress in constraining the Galactic nova rate, but it can still be greatly improved. In Kawash et al.\ (2022, in preparation), we plan to estimate the detection efficiency of ASAS-SN through fake transient recovery and incorporating various decline rates to our simulated sample of novae. Knowing this, along with continued observations from ASAS-SN and PGIR will allow us to further constrain the rate of novae in the Galaxy. We can further quantify the effects of Solar constraint on nova discovery rates and make predictions for next generation transient facilities like LSST.

\section*{Acknowledgments}
AK, LC, EA, and KVS acknowledge financial support of NSF award AST-1751874 and a Cottrell fellowship of the Research Corporation. JS acknowledges support from the Packard Foundation.
BJS, CSK, and KZS are supported by NSF grant AST-1907570. CSK and KZS are supported by NSF grant AST-181440. 

We thank the Las Cumbres Observatory and its staff for its continuing support of the ASAS-SN project. ASAS-SN is supported by the Gordon and Betty Moore Foundation through grant GBMF5490 to the Ohio State University, and NSF grants AST-1515927 and AST-1908570. Development of ASAS-SN has been supported by NSF grant AST-0908816, the Mt. Cuba Astronomical Foundation, the Center for Cosmology and AstroParticle Physics at the Ohio State University, the Chinese Academy of Sciences South America Center for Astronomy (CAS- SACA), and the Villum Foundation. 

The analysis for this work was performed primarily in \texttt{ipython} \citep{pg07} using \texttt{numpy} \citep{oliphant2006guide,van2011numpy}, \texttt{Astropy} \citep{astropy:2018}, \texttt{Matplotlib}  \citep{Hunter:2007}, and \texttt{scipy} \citep{vgo20}.


\bibliographystyle{aasjournal}
\bibliography{biblio}



\appendix
\renewcommand\thefigure{\thesection.\arabic{figure}}    
\setcounter{figure}{0}    

\renewcommand\thetable{\thesection.\arabic{table}}    
\setcounter{table}{0}    
\section{Appendix}

\subsection{Besancon Mass Model}
\label{sec:besancon}
Here we discuss the form of each component of the Besancon Mass model used to distribute novae for our secondary model. The total mass and normalization of each component is shown in Table \ref{table:comp_mass}.

\begin{deluxetable}{lcc}
\tablecolumns{3}
\tablewidth{0pt}
\tablecaption{Mass and Normalization values for the various components of the Galactic model}
\label{table:comp_mass}
\tablehead{
\colhead{Component} & 
\colhead{Normalization} & 
\colhead{Total Mass}\\
\colhead{ } & \colhead{M$_\odot$ pc$^{-3}$} &
\colhead{$10^9$ M$_\odot$} }
\startdata
Thin Disk & 1.45 & 35.0 \\
Thick Disk & 0.002$^{a}$ & 4.67 \\
Bulge (Model S) & 2.37 & 22.1 \\
Bulge (Model E) & 1.17 &  1.20 \\
Halo & 0.00005 &   0.55
\enddata
\tablenotetext{a}{Density at the solar position, where the other normalization values refer to the density at the Galactic center}
\tablecomments{Normalization values and the total mass of the various components of the Galactic model utilized in this work. We set the normalization values to achieve a consistent total mass with that derived in \cite{robin03}, \cite{sbi17}, or \cite{bjg16}}
\end{deluxetable}

We use a Cartesian grid with resolution 0.1 kpc, R = $\sqrt{x^2 + y^2}$ is the radial distance from the Galactic center, and z is the distance perpendicular to the Galactic plane. Throughout the components, the assumed Solar radius from \cite{robin03} is $R_\odot = 8.5$ kpc. 

\subsubsection{Thin Disk}
The form of the thin disk density is from \cite{robin03}

\begin{equation}
    \rho = \rho_0 \times \left\{\exp \left[ - \left(0.5^2 + \frac{a^2}{h_{R_{+}}^2}  \right)\right ] - \exp \left[ - \left(0.5^2 + \frac{a^2}{h_{R_{-}}^2} \right)\right ]  \right \}  
\end{equation}
where $a^2 = R^2 + (z/\epsilon)^2$, $h_{R_{+}}$ = 2.5 kpc is the scale length of the disk, $h_{R_{+}}$ = 0.9 kpc is the scale length of the hole, and $\epsilon$=0.0791. 

\subsubsection{Thick Disk}
A piece-wise thick disk density distribution is utilized from \cite{robin03}
\begin{equation}
    \rho = \left\{\begin{matrix}
\rho_0 \exp \left ( - \frac{R - R_\odot}{h_R} \right ) \times \left(1 - \frac{z^2/h_z}{\xi \times \left(2 + \xi/h_z \right )} \right ) & \rm{if}~ z \leq \xi
\\ 
\rho_0 \exp \left ( - \frac{R - R_\odot}{h_R} \right ) \times \exp \left(- \frac{\left | z - z_\odot \right |}{h_z} \right) \times \frac{2\exp\left(\xi/h_z \right )}{2+\xi/h_z} & \rm{if} ~ z > \xi

\end{matrix}\right.
\end{equation}
where $h_R = 2.5$ kpc is the radial scale length, $h_z = 0.8$ kpc is the vertical scale length, and $\xi = 0.4$ kpc. The local density $\rho_0 = 0.002$ M$_\odot$ pc$^{-3}$ is set to be four percent of the local thin disk density \citep{bjg16}.  

\subsubsection{Bulge/Bar}
For the bulge, we use an updated fit to VVV data from \cite{sbi17}. The best fit model combines a hyperbolic secant density distribution

\begin{equation}
\begin{matrix}
    \rho = \rho_0 \sech^2 \left(r_s\right) & (\rm{model} ~\rm{S})
\end{matrix} 
\end{equation}

and an exponential distribution

\begin{equation}
\begin{matrix}
    \rho = \rho_0 \exp \left( -0.5 r_s^n \right) & (\rm{model}~  \rm{E})
\end{matrix} 
\end{equation}
where, 
\begin{equation}
    r_s = \left \{\left[ \left( \frac{x}{x_0} \right )^{c_\perp} + \left( \frac{y}{y_0} \right )^{c_\perp}\right ]^\frac{c_\parallel}{c_\perp}   + \left( \frac{z}{z_0} \right )^{c_\parallel}\right\}^{1/c_\parallel}.
\end{equation}
$c_\parallel$ and $c_\perp$ controls the face-on and edge-on shape of the bulge, respectively, and $x_0$, $y_0$, and $z_0$ are the scale lengths in each respective direction. We use the best fit parameters using the Besancon discs presented in \cite{sbi17}. For the $\sech$ component (model S) the best fit parameters are $c_\parallel$ = 2.89,  $c_\perp$ = 1.49, and ($x_0$, $y_0$, $z_0$)  = (1.65,0.71,0.50) kpc, and for the exponential component, the best fit parameters are $c_\parallel$ = 3.64,  $c_\perp$ = 3.54, and ($x_0$, $y_0$, $z_0$)  = (1.52,0.24,0.27) kpc, and n = 2.87. The bulge density has a cutoff Radius $R_c = 6.96$ kpc implemented by multiplying the bulge density $\rho$ by the function

\begin{equation}
    \begin{matrix}
    f(R) = 1  & R < R_c \\ 
     f(R) = \exp \left[ -2 \left (R - R_c \right )^2 \right ] & R > R_c.  
    \end{matrix}   
\end{equation}

\subsubsection{Stellar Halo}
We use a power law form of the halo similar to the one presented in \cite{robin03} 

\begin{equation}
\rho = \left\{
    \begin{matrix}
    \rho_0 \left(  a_c/R_\odot\right )^n & a < a_c \\ 
      \rho_0 \left(  a/R_\odot \right )^n & a > a_c  
    \end{matrix}   \right.
\end{equation}
where $a = \sqrt{ x^2 + y^2 + \left(z/\epsilon \right)^2}$, $a_c = 0.5$ kpc is the cutoff radius, and $n = -2.44$.




\label{lastpage}
\end{document}